\documentclass[%
 reprint,
 amsmath,amssymb,
 aps,
 prl,
]{revtex4-1}

\usepackage{graphicx}
\usepackage{dcolumn}
\usepackage{bm}

\begin{document}


\title{Quantum biology on the edge of quantum chaos}

\author{Gabor Vattay}
\author{Stuart Kauffman}%
\affiliation{%
University of Vermont, Vermont Complex Systems Center\\
210 Colchester Ave, Farrell Hall, Burlington, VT 05405}%

\author{Samuli Niiranen}
\affiliation{
 Tampere Institute of Technology,
 Department of Signal Processing\\
P.O.Box 553
FI-33101 Tampere
Finland}%

\date{\today}

\begin{abstract}

We give a new explanation for why some biological systems can stay quantum coherent for long timeS at room temperatures, one of the fundamental puzzles of quantum biology. We show that systems with the right level of complexity between chaos and regularity can increase their coherence time BY orders of magnitude. Systems near Critical Quantum Chaos or Metal-Insulator Transition (MIT) can have long coherence times and coherent transport at the same time. The new theory tested in a realistic light harvesting system model can reproduce the scaling of critical fluctuations reported in recent experiments. Scaling of return probability in the FMO light harvesting complex shows the signs of universal return probability decay observed at critical  MIT. The results may open up new possibilities to design low loss energy and information transport systems in this Poised Realm hovering reversibly between quantum coherence and classicality. 

\end{abstract}


\maketitle


Discovery of room temperature quantum coherence in the avian compass\cite{PhysRevLett.106.040503} of birds, in the olfactory receptors\cite{Turin} 
and in light harvesting complexes\cite{Nature.10.1038,Nature08811,Panitchayangkoon20072010,PNAS2011} in the last few years indicate that 
quantum effects might be ubiquitous in biological systems. While the quantum chemical understanding of the details of light harvesting systems is 
almost complete, no organizing principle has  been found which could explain why quantum coherence is maintained in these systems for much longer than the
characteristic decoherence time imposed by their room temperature environment.
Here we propose that at the critical edge of quantum chaos coherence and transport can coexist
for several orders of magnitudes longer than in simple quantum systems. Quantum systems changing from integrable to
quantum chaotic pass through critical quantum chaos\cite{ Altshuler, Efetov, PhysRevB.47.11487, Evangelou} which is also a metal-insulator
transition from Anderson localization to extended wave functions. By extending the semiclassical theory of decoherence
from chaotic\cite{Zurek1995300, PhysRevLett.70.1187, PhysRevLett.83.4526, PhysRevLett.90.014103, PhysRevE.56.5174, PhysRevLett.77.59, PhysRevE.60.1643}
and integrable systems\cite{PhysRevLett.92.150403} to the transition region we show that coherence decay changes from
exponential to power law behavior and coherence time is amplified exponentially from its environmentally determined value. 
We demonstrate our theory on a ring of chromophores passing through the critical point and show that
coherence in the critical point decays with the same non-trivial power law as in the FMO complex experiment\cite{Panitchayangkoon20072010}.
Our results also suggest that loss of coherence is not permanent in these systems and they can re-cohere via  
coherent external driving such as the arrival of photons in case of the light harvesting systems and can continuously hover in the "Poised Realm"\cite{Kauffman}
between the coherent quantum and the incoherent classical worlds. Using this new critical design principle from biology might open the
way to build lossless quantum coherent energy and information processing devices operating at room temperature.
\begin{figure}[ht]
\centerline{\includegraphics[width=0.5\textwidth]{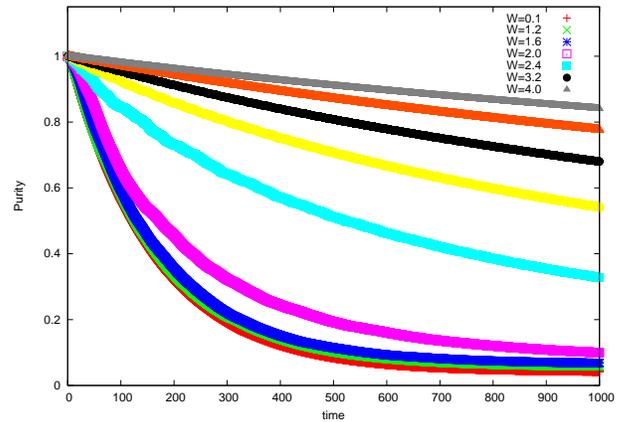}}
\caption{Purity decay of the chromophore ring with 1D Harper hamiltonian. Below the metal-insulator transition $W<W_c=2$
curves can be fitted with exponentials $P(t)=\exp(-t/\tau_c)$. In the parameter
range $W=0.1-1.9$ the fitted coherence time changes in the range $\tau_c=182-248$.
In and above the transition $W\geq W_c$ the curves can be fitted with $P(t)=(\tau_\alpha/(\tau_\alpha+t))^\alpha$.
In the parameter range $W=2.0-4.0$ the exponent changes in the range $\alpha=2.01-0.479$
and $ \tau_\alpha=451-1651$. The estimate for the decoherence time $P(\tau_{c})=1/e=37\%$ 
above the transition is $\tau_{c}=\tau_{\alpha}(e^{1/\alpha}-1)=266-16051$.}
\end{figure}

Quantum biology is dealing with open quantum systems closely coupled to their many degrees of freedom environment. The environment 
exerts time dependent forces on the system through the coupling. Some of these forces change very rapidly compared to the excitation frequencies
of the system and look random from its point of view. This "heat bath" destroys  quantum coherence and moves the system into a mixed state rapidly. 
The average effect of the random forces can be described as a non-unitary time evolution of the system's density matrix.

At room temperature the phonon environment  has characteristic energy $E_T=k_BT\approx 4\cdot{10^{-21}}J$ and frequency $\nu_T=k_BT/h\approx 6 THz$ 
in the infrared spectrum.  The typical phonon thermal wavelength is determined by the 
speed of sound of the material the system is embedded in. Its typical value for water and non-specific proteins is $c=1500 m/s$ 
yielding $\lambda_T=c/ \nu \sim 2.5\AA$. 
Molecules within a thermal wavelength distance can feel the same environment and some of their states can be almost decoherence free\cite{Lidar} and can preserve coherence for long time. In light harvesting systems the light absorbing and
emitting chromophores are embedded in a protein scaffold which can suppress  thermal fluctuations more effectively. The thermal wavelength can be increased to $10-13\AA$ by raising the effective speed of sound \cite{Nalbach} with a factor of 3-5 to $7-8000m/s$.
Protective environments cannot extend the size of coherent patches further and a different mechanism is needed to
extend it further, which we propose next.
\begin{figure}[ht]
\centerline{\includegraphics[width=0.5\textwidth]{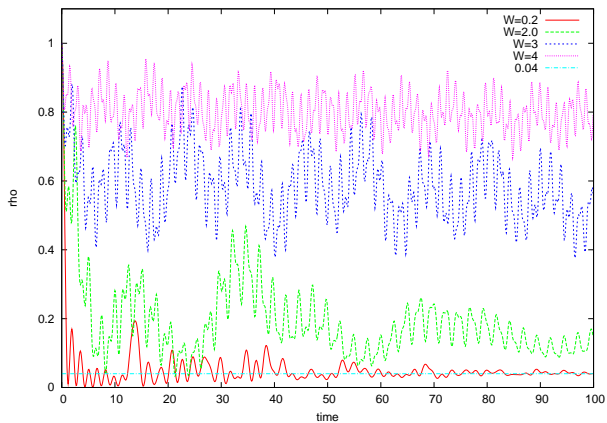}}
\caption{Probability that the exciton stays on the chromophore it started in represented by the density matrix element 
$\varrho_{1,1}$. Below the transition ($W=0.1$) coherence dies out quickly and the probability reaches its asymptotic value $1/25$. The peak at  $time\approx13$ is the result of the interference of waves going clock and anti-clockwise  along the the circle and meeting 
again after turning around the structure. The rest of the structure comes from interference of waves scattering back from other chromophores. In the transition point and above coherent beats occur and the probability stays elevated for a very long time (not shown here).
These are beats due to genuine quantum coherent superposition states .  }
\end{figure}

The speed of  environmental decoherence can be  characterized by the decay rate of the off diagonal ($n\neq m$) elements of the reduced density matrix 
of the system $\varrho_{nm}\sim e^{-t/\tau_c},$ where $\tau_c$ is the coherence time. Purity $P=Tr[\varrho^2] =\sum_{mn} |\varrho_{nm}|^2$ has been shown to 
be a good overall measure. It is $P=1$ when the system is in a pure state and decreases monotonically as the system decoheres into a mixed state.
$P(t)\sim 1/N+ const \cdot e^{-t/\tau_c}$, where $N$ is the number of quantum states  of  the system. The logarithm of the purity is the Renyi entropy $S_2=\log Tr[\varrho^2] $
 of the system. The long time entropy production rate  of the system\cite{PhysRevLett.70.1187} and the rate of decoherence are then closely related $dS_{2}/dt \sim  1/\tau_c$ for $t\rightarrow\infty$.
 Entropy production on the other hand is determined by the dynamical properties of the system. It has been derived via 
semiclassical approximation and then proven by direct simulations that the  entropy production rate becomes environment independent and is determined by the  
classical dynamical Kolmogorov-Sinai entropy
of the system\cite{Zurek1995300,PhysRevLett.70.1187,PhysRevLett.83.4526,PhysRevLett.90.014103,PhysRevE.56.5174,PhysRevLett.77.59,PhysRevE.60.1643}.
It is in turn the sum of the positive Lyapunov exponents $\lambda_i^+$  characterizing the exponential divergence of chaotic 
trajectories in the system $dS_2/dt \sim h_{KS}=\sum_i \lambda^+_i.$ This relation between coherence decay and generalized Lyapunov exponents has been confirmed in 
strongly chaotic systems.  Another implication of this result is that the rate of decoherence vanishes in systems where the Lyapunov exponent is zero.
This has also been confirmed in integrable systems. These are completely solvable systems with fully predictable regular dynamics and  
zero Lyapunov exponents. Purity shows power law decay typically like $P(t)\sim 1/t^2$ and asymptotic decoherence rate is zero $dS_2/dt\sim 1/t \rightarrow 0$.

Zero Lyapunov exponent and entropy production can also emerge in systems at  the border of the onset of global chaos in the classical counterpart of the system. 
Suppose, we have a parameter $\epsilon$ of the mechanical system which characterizes its transition from regular dynamics to chaos\cite{lichtenberg1992regular, reichl2004transition}
$H=H_R+\epsilon H_C,$ where $H_R$ is the Hamiltonian of a fully integrable system and $H_C$ is fully chaotic. 
Classically $H_R$ is a solvable system and it can be described by action-angle variables.  It does only simple oscillations in the angle variables while the action variables do not
change and remain conserved restricting the dynamics for the surface of a torus in the phase-space. 
When $\epsilon\neq0$ but small the system is not integrable classically and the  Kolmogorov-Arnold-Moser (KAM) theory describes the system\cite{lichtenberg1992regular, reichl2004transition}. 
The chaotic perturbation breaks up some of the regular tori in the phase-space and chaotic diffusion emerges localized between unbroken, so called KAM tori.  Chaotic regions are
localized in small patches in the phase-space surrounded by regular boundaries represented by the KAM tori. At a given critical $\epsilon_c$ even the last KAM tori separating the system gets broken 
and the chaotic patches merge into  a single extended chaotic sea.   In the transition region $\epsilon\approx\epsilon_c$ the Lyapunov exponent shows a 
second order phase transition\cite{mackay1993renormalisation} with power law scaling $\lambda_0(\epsilon)\sim (\epsilon-\epsilon_c)^{\beta}$ slightly above $\epsilon>\epsilon_c$ with some exponent $\beta>0$.
Above the transition $\epsilon>\epsilon_c$ the system is chaotic characterized by a positive largest Lyapunov exponent $\lambda_0>0$. 

On the quantum mechanical level we can follow the transition in the statistical distribution of the energy levels. The Hamilton operator of the regular system $H_R$ 
is a separable with diagonal matrix elements.  The consecutive energy levels of the regular system look random and follow a Poisson process.  The nearest neighbor level spacing distribution is 
then exponential $p(s)=\exp(-s),$ where $s_n=(E_{n+1}-E_n)/\Delta(E_n)$ is the level spacing measured in the units of local mean level spacing $\Delta(E)$ at energy $E$.
The Hamiltonian operator $H_C$ corresponding to the fully chaotic system is best approximated by a random matrix. The energy level statistics of $H_C$ can be described by Random Matrix Theory (RMT) 
and the level spacings follow the Wigner level spacing distribution\cite{PhysRevLett.52.1} $p(s)={(\pi s}/2) \exp(-\pi s^2/4)$ in systems with time reversal symmetry.
As the parameter $\epsilon$ is increased from zero the level spacing statistics changes from a Poissonian to a Wigner distribution. Critical quantum chaos\cite{Evangelou}  sets up at the critical value $\epsilon_c$ in between.  
Below the critical point $p(0)$ is finite, at the critical point and above the spacing distribution starts linearly $p(s)=As$ for $s\rightarrow0$, a characteristic feature of chaotic systems with strongly overlapping eigenfunctions. 
The tail of the distribution remains exponential below the critical point  $\exp(-Bs)$ for $s\rightarrow\infty$ which is a characteristics of regular systems whose eigenfunctions do not overlap for larger energy separations.
It turns to gaussian $\exp(-Cs^2)$ then above the critical point. At the critical point the level statistics is semi-Poisson\cite{Evangelou} $p(s)=4s\exp(-2s)$ which starts linearly and
decays exponentially combining the two main aspects of the level statistics of regular and chaotic systems. 
\begin{equation}
\partial_t\varrho_{nm}=\frac{1}{i\hbar} [ H,\varrho]_{nm} - \frac{1}{\hbar^2}(C_{nn}+C_{mm}-2C_{nm})\varrho_{nm},\label{lindblad}
\end{equation} 
where $C_{nm}=\langle F(x_n,t)F(x_m,t)\rangle$ is the correlation function of the environmental coupling. We assume that the correlation function depends
only on the periodic distance of the chromophores in a simple way $C_{nm}=D(L/\pi)^2\cos^2(\pi(n-m)/L)$ and is quadratic for small distances.

The transition described here is more general than just the transition from regular to quantum chaotic behavior. It is also a transition from the localized states of the regular system
to the extended states of the chaotic system. The two are separated by the metal-insulator transition (MIT) point\cite{Altshuler,PhysRevB.47.11487} between quantum mechanical Anderson 
localization and globally delocalized metallic phase. The transition point can be identified with the emergence of the semi-Poissonian\cite{PhysRevB.47.11487} level statistics.
In the transition point the wave functions are neither fully localized nor extended and have an intriguing multi-fractal spatial character. The fractal structure allows them to develop a hairy localized 
structure but also an extended structure with long range overlap correlations.

Merging the pieces of classical, semiclassical and quantum aspects a new picture emerges. Systems well below the critical point have non-chaotic dynamics
with zero generalized Lyapunov exponents and quantum localization lengths extending only for few states. Decoherence in these systems is slow and purity follows a power law decay
$P(t)\approx 1/t^\alpha$ with some exponent $\alpha$ making possible the presence of anomalously long living coherent dynamics in the system. But coherently evolving states remain 
localized and long distance quantum coherent transport is not possible. Systems  well above the critical point have chaotic dynamics with positive Lyapunov exponents and delocalized states extending for the entire system.
Coherence dies out exponentially fast. Near the critical point exponential decay of coherence crosses over to long living power law behavior and localized states
become delocalized. In finite systems there is always a narrow region around criticality, where long living coherence and sufficiently extended states can exist at the same time.

\begin{figure}[ht]
\centerline{\includegraphics[width=0.5\textwidth]{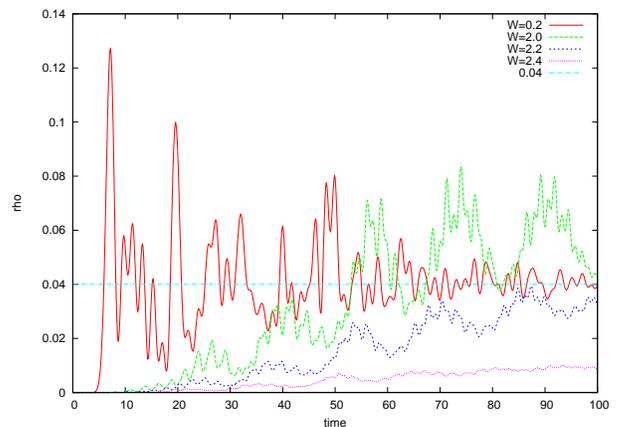}}
\caption{Probability that the exciton started on chromophore 1 is on chromophore 13 $\varrho_{1,13}$. 
Below the transition  ($W=0.1$) coherence dies out quickly and the probability reaches its asymptotic value $1/25$.
The beats at times $6.5$ and at $19.5$ reflect interference from clock and anti-clockwise traveling waves interfering
after taking  half and 1.5 rounds. In the transition point and above we can see that due to the localization it takes much longer time 
for the excitation to arrive at the opposite end. If we are just slightly above the transition coherent beats smear out
by the time they arrive. The reason is that propagation becomes mostly classical as localization stops quantum diffusion.  
In the critical point however  quantum propagation and coherence is possible at the same time.   }
\end{figure}

We demonstrate this theory on a simplified model of chromophores in light harvesting complexes and argue  that it is very likely that biological systems use this mechanism to
tune their parameters\cite{Lloyd} near the critical point to maintain rich quantum transport properties. 
The excitonic states are described in the single excitation approximation by the Hamiltonian $H_{ij}=\sum_i E_i |i \rangle \langle i |+\sum_{ij}V_{ij}|i\rangle\langle j|$,
where $|i\rangle$ indexes the excitonic states with site energies $E_i$ and dipole interaction strengths $V_{ij}$. For simplicity we take a simple ring of  $L$ chromophores coupled
by constant $V_{nm}=1$ for neighboring sites $n$ and $m$ and zero otherwise and take quasi random on site energies given by $E_n=W\cos(2\pi\sigma n)$, where the irrational number $\sigma=(\sqrt{5}-1)/2$ is
the golden mean. This hamiltonian
is known as the one dimensional Harper model. At  $W_c=2$ the infinite system $L\rightarrow\infty$ goes through a MIT with delocalized states below and localized states above criticality.
At the critical point it has been shown to have semi-Poisson level statistics\cite{PhysRevLett.84.1643}. The system is coupled to the phononic environment 
via the Hamiltonian $\sum_i F(x_i,t) |i\rangle\langle i |$, where $F(x,t)$ is the randomly fluctuating phonon field including the chromophore site energy coupling constant.
The reduced density matrix of the chromophore system can be described in Markovian approximation by the Lindblad equation\cite{springerlink:10.1007/BF01608499}

Next we show results for $L=25$ (in dimensionless units $\hbar=1$), which is a realistic number of chromophores in experimentally investigated systems\cite{Olson}. 
In Fig. 1 we show purity of the system. Below the critical point purity decays exponentially. At and above the critical point the curves can be fitted with power law 
exponents changing from $\alpha\approx2$ at criticality towards zero as $W$ increases and the curves flatten out. In Fig. 2 we show the probability $\varrho_{1,1}$ to find the exciton on the chromophore in which 
the exciton was initialized. Below criticality the probability reaches its asymptotic value of $1/L=0.04$ quickly after decaying coherent oscillations. Above criticality the probability
stays above the asymptotic value for a long time indicating the presence of localization. \begin{figure}[ht]
\centerline{\includegraphics[width=0.5\textwidth]{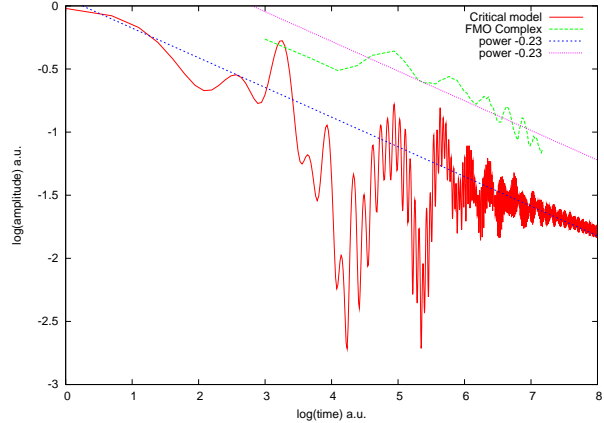}}
\caption{  Power law decay of the population probability $\varrho_{1,1}$ for the chromophore model and for the real FMO complex.
The population decay for a modified version of our model is shown (red) on log-log scale at the critical point $W=2$, where each 
site is fully correlated with its next neighbors and
not correlated with the rest of the chromophores.  The same quantity is shown for the FMO complex (green).  Time scales are in
arbitrary units. Both curves oscillate around a trend which decays with the same exponent of about $-0.23$. While the geometry
is different for FMO and the ring, both of them seem to follow the same, perhaps universal, scale free trend observed at the critical point of the MIT. 
 }
\end{figure}Quantum beats can be observed which also relax in a very slow fashion.
Based on the simulations we can establish the rule of localization assisted amplification of coherence time. In the delocalized regime purity decays exponentially determined
by the timescale dictated by the environment. In the localized regime we can define an effective coherence time by looking at the point where purity decays to $1/e$ of its original value $P(\tau_{c})=1/e$. 
Above the critical point purity can be well approximated by function $P(t)=(\tau_\alpha/(\tau_\alpha+t))^\alpha$ (see Fig. 1). The effective length of coherence then can be approximated as $\tau_{c}=\tau_{\alpha}\cdot(e^{1/\alpha}-1)$.  
This function grows very fast when $\alpha\rightarrow 0$ in the strongly localized limit. In our example this is a 60-fold increase between criticality and $W=4$. In Fig. 3 we show the probability $\varrho_{13,13}$ of finding the excitation at the opposite end of the ring. For subcritical values the excitation arrives very quickly due to delocalization and shows
beats due to the interference of excitons traveling clock and anti-clockwise. Coherent beats die out quickly and we reach the asymptotic probability. For supra-critical
values far from the critical point the probability to reach site 13 remains astonishingly low due to the localization of the system. For values near at and below criticality we get the 
most optimal results for quantum coherent transport of excitations, when excitations can still reach the opposite end of the circle but can preserve a degree of coherence as well.

At criticality not only purity changes from exponential to power law decay but so does the population of the chromophores. In Fig. 4 we show the population $\varrho_{1,1}(t)$
in a version of our model where three neighboring chromophores  along the circle are always fully correlated $C_{nm}=C$ while they become totally uncorrelated otherwise $C_{mm}=C\rightarrow\infty$.
This model can describe the real situation where neighboring chromophores are shielded from the environment by their protein scaffolds and approximately three chromophores
can be placed within the protected thermal wavelength of 10-13$\AA$ . We can see that the
trend of the population follows a shallow power law decay. We show also the experimental data of Ref. 5 on the FMO light harvesting complex kindly provided by the authors. Both curves follow a similar
scale free trend with approximately the same exponent $-0.23$. 
This exponent is very close to $-0.25$ which is the power law decay exponent of the average return
probability $p(t)=\langle p_n(t)\rangle_n\sim t^{-1/4}$ at the critical point of MIT  as it was shown in Ref.\cite{Katsanos}.
The return probability $p_n(t)=\langle \mid \psi_n(t)\mid^2\rangle $ is the probability of return assuming that the wave function was localized on the site initially $\psi_n(0)=1$. In the decoherence free case
it coincides with the density matrix element $\varrho_{nn}(t)$ assuming $\varrho_{nn}(0)=1$, which is
shown in our model and for the FMO complex. It seems likely that the FMO complex follows the
universal scaling of critical MIT indicating that  the hamiltonian of the FMO complex is tuned to critical quantum chaos in order to realize optimal coherent transport, what we show elsewhere.

The findings support a new approach to quantum biological systems. They are not just under the influence of environmental decoherence due to random noise but also driven
by the coherent waves of the incoming photons. The photons are absorbed by one of the chromophores which can be interpreted as a measurement process selecting
one of the chromophores randomly. Then the system is set into an initial state which is concentrated on the selected chromophore. The purity of the system becomes $P=1$
as this is a pure state and the partially decoherent evolution starts again decreasing the purity in time. The system can hover in the "Poised Realm"\cite{Kauffman} between clean quantum and 
incoherent classical worlds. By tuning the timings of re-coherence events and the coherence time during decoherence via tuning the system on the chaos-regularity axis can be kept in high 
level of purity. 

We hope that the mechanism discovered here makes  it possible to create new quantum devices
working at room temperature tuned to critical MIT capable of nearly frictionless quantum transport of energy and information. 


The authors thank for the financial support of Lockheed Martin Co.. S.A.K. thanks the support of the TEKES Foundation, Finland.

\end{document}